\documentclass[b5paper,10pt,abstractoff,DIV=calc,headings=normal,twoside]{scrartcl}
\usepackage[english]{babel}
\usepackage{lmodern}
\usepackage{microtype}
\usepackage{hyperref}
\usepackage[hyphenbreaks]{breakurl}
\usepackage{amsmath,amssymb,amsthm,amsfonts,graphicx,etoolbox,scrlayer-scrpage}
\usepackage[noblocks]{authblk}
\makeatletter
\patchcmd{\@maketitle}{\huge}{\Large}{}{}
\patchcmd{\abstract}{\quotation}{}{}{}
\AtBeginEnvironment{abstract}{\noindent\ignorespaces}
\AtEndEnvironment{abstract}{\par\mbox{}}
\newcommand{\shortauthor}{}
\newcommand{\shorttitle}{\@title}
\makeatother
\setkomafont{author}{\small}
\setkomafont{title}{\rmfamily\bfseries}
\setkomafont{disposition}{\rmfamily\bfseries}

\def\AMS#1{\par\noindent \textbf{AMS subject classification: }#1\par}

\newcommand{\keywords}[1]{\par\noindent\textbf{Keywords: }#1}
\rehead[]{\shortauthor}\lohead[]{\shorttitle}\lehead[]{PPP}\rohead[]{PPP}
\ofoot[]{}\ifoot[]{}\recalctypearea
\theoremstyle{plain}

\theoremstyle{definition}

\theoremstyle{remark}

\renewenvironment{abstract}{\bigskip\noindent\begin{minipage}{\textwidth}\setlength{\parindent}{15pt}\paragraph{Abstract:}}{\end{minipage}}

\usepackage{bm}
\usepackage{hyperref}

\begin{document}


\renewcommand{\shortauthor}{J. Lachmann and A. Hubin}

\title{Subsampling MCMC for Bayesian
Variable Selection and Model Averaging in
BGNLM}

\author[1]{Jon Lachmann\thanks{Corresponding author: jon@lachmann.nu}}
\affil[1]{Stockholm University, Indicio Technologies AB}
\author[2]{Aliaksandr Hubin}
\affil[2]{NMBU, University of Oslo, OUC} 
\date{}
\maketitle
\begin{abstract}
    Bayesian Generalized Nonlinear Models (BGNLM) offer a flexible nonlinear alternative to GLM while still providing better interpretability than machine learning techniques such as neural networks. In BGNLM, the methods of Bayesian Variable Selection and Model Averaging are applied in an extended GLM setting. Models are fitted to data using MCMC within a genetic framework by an algorithm called GMJMCMC. In this paper, we combine GMJMCMC with a novel algorithm called S-IRLS-SGD for estimating the marginal likelihoods in BGLM/BGNLM by subsampling from the data. This allows to apply GMJMCMC to tall data.
\end{abstract}
\smallskip
\AMS{62-02, 62-09, 62F07, 92D20, 90C27, 90C59} 

\section{Introduction}\label{sec:intro}

In \cite{hubin2020flexible}, the class of \textit{Bayesian Generalized Nonlinear Models (BGNLM)} is proposed to provide flexible (potentially deep Bayesian) modelling of data with high interpretability. More specifically, BGNLM models the relationship between one dependent variable $ \bm y = (y_1,...,y_n)$ and $ m $ independent variables $ \bm x = (\bm x_1,...,\bm x_n)^T$, where $ \bm x_i = (x_{i1},..., x_{im}) $ with observation count $n$. This is done through an extended \textit{Generalized Linear Model (GLM)}. In a general sense, the class of models is defined as
\begin{align}
	y | \mu, \phi &\sim \mathfrak{f}(y | \mu, \phi), \label{dens1}\\
	\mathsf{h}(\mu) &= \beta_0 + \sum_{j=1}^{q} \gamma_j \beta_j F_j(\bm x, \bm \alpha_j)\label{bgnlm1}.
\end{align}
In Equation~\eqref{dens1}, $ \mathfrak{f}(\cdot | \mu, \phi) $ is a density/mass function of a distribution from the exponential family with mean and dispersion parameters $ \mu $ and $ \phi $ respectively. In \eqref{bgnlm1}, $ \mathsf{h}(\mu) $ is an link function between the mean $ \mu $ and the \textit{features} $ F_j $. The features are nonlinear transformations of covariates, the definition of which is discussed in detail in \cite{hubin2020flexible}. For now, we will just note that a single feature can contain both bare covariates and other features, resulting in a recursive structure that allows for high flexibility. In Equation~\eqref{bgnlm1}, each feature $ F_j(\bm x, \bm \alpha_j)$ has two \textit{outer} parameters: The binary parameter $ \gamma_j $ which is 1 if feature $ F_j $ is included, and 0 otherwise and the coefficients $ \beta_j $ which works as in a regular linear model, specifying the influence of feature $ F_j $ in this particular model. Each feature also has a number of \textit{inner} parameters $ \bm \alpha_j $, that define its structure. Moreover, by introducing some hard and soft constraints on feature complexity that are defined in \cite{hubin2020flexible}, the set of $ q $ possible features $ \mathcal{F} $ is forced to be finite.

To describe the feature space $ \mathcal{F} $ of BGNLM, the following properties of each feature are introduced in \cite{hubin2020flexible}: \textit{depth}, \textit{width} and \textit{operations count}. In \cite{hubin2020flexible} they are used to define three hard constraints which are applied to limit the model space and make sure that it is finite. These constraints are: The maximum depth of a feature, $ D $; The maximum width of a feature, $ L $; The maximum total number of features in a given model $ Q $.

Further, in \cite{hubin2020flexible}, a model prior which penalises more complex models is defined. Each model is identified by the vector $ \bm \gamma $, indicating the features that it involves, i.e. $ \mathfrak{m} = (\gamma_1,...,\gamma_q) $. Using some parsimonious complexity measure $ c(F_j(\cdot, \bm \alpha_j)) \geq 0 $ of a feature $ F_j(\cdot, \bm \alpha_j) $, the model prior is then defined as
\begin{align}
	p(\mathfrak{m}) \propto \prod_{j=1}^{q} u^{\gamma_j c(F_j(\cdot, \bm \alpha_j))},
\end{align}
Here, $ u \in (0,1) $ will give us a larger negative number for models including more numerous and complex features. This will cause the less complicated of two models that are equally good at describing the data to be favoured. We shall follow \cite{hubin2020flexible} and use the \textit{operations count ($oc$)} of a feature as the complexity measure. It simply counts the number of algebraic manipulations ($+$,$-$,$*$, and nonlinear transformations) in the feature: $x_1$ has $oc$ of 0 but $sin(x_1*x_2)$ of 2.

\section{Genetically Modified MJMCMC (GMJMCMC)}

MCMC or MJMCMC \cite{hubin2018mjmcmc} working on the marginal space of models cannot be directly applied to explore the feature space from Equation~\eqref{bgnlm1}. First, the model space of size $2^q$ increases exponentially with the number of features $q$. Second, $q$ is growing exponentially with the depth of the features. Hence, an alternative solution is required. In \cite{hubin2020logic}, this problem is solved in the context of Bayesian logic regressions by introducing an algorithm called \textit{Genetically Modified MJMCMC (GMJMCMC)}, which adaptively embeds MJMCMC into a genetic programming framework. It was further adopted in \cite{hubin2020flexible} for BGNLM.

Beginning with an initial population of features $\mathcal{S}_{0}$ consisting of the bare covariates, MJMCMC is applied to explore the model space spanned by this population. A new population is then generated by randomly filtering features that have low marginal inclusion probability. Then, replacements are generated by applying the transformations corresponding to the feature generating process \cite{hubin2020flexible} on the remaining features. This process is repeated a set number of times $T$, referred to as the number of populations of that particular run. Further, in \cite{hubin2020logic,hubin2020flexible} embarrassing parallelisation of the algorithm is suggested. Also, recurrence of the algorithm is shown under practical assumptions. 

\paragraph{Proposed subsampling technique}
 Markov Chain Monte Carlo (MCMC) algorithms such as Metropolis-Hastings are very useful to sample from complex posterior distributions, but as the amount of data involved increases, so does computational time. To address this issue subsampling MCMC methods start to appear. Since our interest in this paper is within Bayesian model selection and model averaging, a specifically interesting for us paper is \cite{lachmann2022subsampling}, which develops and confirms numerically and theoretically a subsampling-based stochastic optimisation technique for computing the marginal likelihoods and is coupled with MCMC. In \cite{lachmann2022subsampling}, it is proven that any converging stochastic optimisation can be used in theory, yielding a time inhomogeneous MCMC (TIMCMC) with the desired limiting distribution, yet recommend a stochastic gradient descent with subsampling iterated reweighted least squares initialisation (S-IRLS-SDG) as a working practical solution. S-IRLS-SDG effectively uses a subsampling IRLS algorithm for fast convergence to the proximity of the local optima, which is further obtained by a standard SGD approach. Since GMJMCMC is a recurrent algorithm, the convergence result from \cite{lachmann2022subsampling} automatically extends to it.

In this paper, we suggest using S-IRLS-SDG available at \url{https://github.com/jonlachmann/irls.sgd} as the TIMCMC within the populations of GMJMCMC resulting in a new subsampling-based GMJMCMC algorithm. Our implementation is available as an R-package at \url{https://github.com/jonlachmann/GMJMCMC}. Further, we provide experiments showing this to be a working solution for GMJMCMC and BGNLMs.

\section{Experiments}

\paragraph{Reproducibility study}
As an example to demonstrate the possibility to use BGNLM for inference, in \cite{hubin2020flexible} it was applied to a subset of the Open Exoplanet Catalogue data \cite{rein2012proposal}. This dataset contains information about planets that are not in our solar system. The dataset contains 11 variables, outlined in \cite{hubin2020flexible}. 

Kepler's third law \cite{kepler1609} relates the square of a planet's orbital period $ P^2 $ to the cube of the semi-major axis of its orbit $ a^3 $. Kepler's original formulation of the law corresponds to
$
	a \approx K(P^2 M_h)^{1/3}, 
$
where  $ M_h $ is the mass of the star the planet is orbiting and the gravitational constant $ G $ and $ 4 \pi^2 $ are replaced with the constant $ K $. By setting $ a $ to be the dependent variable and the remaining variables as independent, \cite{hubin2020flexible} were able to apply BGNLM to recover the law. Noting that the correlation between $(P^2 M_h)^{1/3}$, $(P^2 R_h)^{1/3}$ and $(P^2 T_h)^{1/3}$ (with $R_h$ being the radius of the star and $T_h$ its temperature)
 was almost 1, all these three variants were considered as true positives (TP).

Following \cite{hubin2020flexible}, we consider every feature with a posterior marginal probability above $ 0.25 $ to be a discovered feature. Every time we merge multiple runs, we count how many of the discovered features were \textit{not} any of the ones in defined true positives, and this count is considered \textit{false positives (FP)}. For the merged results that \textit{did} include at least one of the true features with a posterior marginal probability above $ 0.25 $, we consider this merged result to contain exactly one \textit{true positive (TP)}. The final measure, \textit{false discovery rate (FDR)} is the proportion of false positives in the total discovered features. From these definitions, and $K$ runs, we calculate the \textit{Power}, the expected number of false positives and FDR as in \cite{hubin2020flexible}.

To demonstrate the efficiency of the new implementation of GMJMCMC, we replicated the results from \cite{hubin2020flexible} in this experiment. We used the same set of nonlinear transformations and  parameters as in the original experiment. Using the measures mentioned above, the results are presented in the left panel of Table~\ref{Tab_a_b}. We see that for 64-thread runs we  recover the true feature 95\% of the time. As expected, the power is lower when fewer threads are used. Our results are almost on par with those presented in \cite{hubin2020flexible}, yet there are still some tuning parameters that might need to be adjusted properly.

\begin{table*}
\caption{\footnotesize Power and false discovery rate (FDR) for}\label{Tab_a_b}
\centering
\footnotesize
\begin{tabular}{c c}

Kepler's third law & Simulated tall data  \\ 
\centering
{\begin{tabular}{lll}
\hline

		Threads & Power & FDR \\
		\hline
		1 & 0.06 (0.14) & 0.97 (0.86) \\
		16 & 0.56 (0.84) &  0.69 (0.18) \\
		64 & 0.95 (1.00) &  0.13 (0.01) \\\hline
\end{tabular}}
&
 {  \begin{tabular}{lll}
\hline
		Threads & Power & FDR \\
		\hline
		1 & 0.42 & 0.89 \\
		32 & 0.67 &  0.78 \\
		128 & 0.76 &  0.74 \\ \hline
\end{tabular}}

\end{tabular}
\end{table*}

\paragraph{Simulation study with tall data ($n \gg m$)}

Next, to check how subsampling influences the convergence of GMJMCMC, we test how GMJMCMC equipped with S-IRLS-SGD performs in terms of recovering the true features from the data generative process. Example 7 in the appendix of \cite{hubin2020flexible} used a set of data with complex interactions. It consisted of 50 binary covariates $ \bm x \sim \text{Bernoulli}(0.5) $ and one dependent variable $ \bm y \sim \text{N}(\bm x^T \bm \beta, 1) $, with $n=1,000$. The coefficient vector was set to $ \bm \beta = (1,1.5,1.5,6.6,3.5,9,7,7,7) $ for the data generative features and $0$ otherwise and the expected value of $ \bm y $ was specified as $E(\bm y | \bm x, \bm \beta)  = \bm \beta_0 + \bm \beta_1 \bm x_7 + \bm \beta_2 \bm x_8 + \bm \beta_3 \bm x_{18}\bm x_{21} +  \bm \beta_4 \bm x_2 \bm x_9  + \bm \beta_5 \bm x_{12} \bm x_{20} \bm x_{37} +  \bm \beta_6 \bm x_1 \bm x_3 \bm x_{27}
    + \bm \beta_7 \bm x_4 \bm x_{10} \bm x_{17} \bm x_{30} + \bm \beta_8 \bm x_{11} \bm x_{13} \bm x_{19} \bm x_{50}.$

To be able to see the performance of our implementation of GMJMCMC combined with S-IRLS-SGD, we generated a tall data set $n=100,000$, and the signal strength adjusted to be equivalent to those from \cite{hubin2020flexible} by increasing the variance of the observations.

We chose to use $ 0.75\% $ of the observations per S-IRLS-SGD iteration and ran GMJMCMC on the data with $ n=100,000 $. For every population, MJMCMC was allowed to explore the model space for 1000 iterations.  All three runs were repeated for 100 times on 1, 32, and 128 threads. In the right panel of Table~\ref{Tab_a_b}, the results from the experiment with $n=100,000$ observations using S-IRLS-SGD with 0.75\% of the observations at each iteration and $1,000$ MJMCMC iterations per population are presented. We observe improvements in both Power and FDR as the computational effort increases. 

\paragraph{Abalone shell predictions}
The Abalone dataset, available at \url{https://archive.ics.uci.edu/ml/datasets/Abalone}, is a long-standing reference for predicting abalone age based on physical measurements. Following \cite{hubin2020flexible}, we divided 4177 observations into a set of 3177 for training and 1000 for testing.

We chose to use $ 10\% $ of the observations per S-IRLS-SGD iteration. We ran GMJMCMC with 32 threads on the Abalone data with other tuning parameters equivalent to \cite{hubin2020flexible} and $u = \exp(-2)$ corresponding to the AIC penalty. The results are summarised in Table \ref{Tab_preds} and show almost identical performance as compared to the full sample analysis. 

\begin{table}[t]{
\caption{\label{Tab_preds} \footnotesize Prediction performances in terms of Root Mean Square Error (RMSE), Mean Absolute Error (MAE), and Pearson's correlation  (CORR).} 
\centering
\footnotesize
\begin{tabular}{llll}%
\hline 
Model&RMSE&MAE&CORR\\\hline
BGNLM\_SUB&1.9506 (1.9343,2.0121)&1.4443 (1.4239,1.4653) &0.7854 (0.7738,0.7892)\\
BGNLM&1.9573 (1.9334,1.9903)&1.4467 (1.4221,1.4750)&0.7831 (0.7740,0.7895)\\
\hline
\end{tabular}}
\end{table}

\section{Conclusion}

We developed a new implementation of GMJMCMC empowered with subsampling through S-IRLS-SGD. The resulting approach is available as an R package \url{https://github.com/jonlachmann/GMJMCMC}. Its appearance on \textit{CRAN} will hopefully help to spark interest in BGNLM and GMJMCMC to get more researchers involved in developing the techniques further even for tall data sets. The reduced computational complexity (detailed time comparisons are available in \cite{lachmann2022subsampling}) enables larger sets of data for BGNLM at the cost of no or little inferential and predictive performance as shown in the examples. %






\bibliography{references}

\end{document}